\documentclass[11pt]{article}

\usepackage[T1]{fontenc}
\usepackage[utf8]{inputenc}
\usepackage{lmodern}
\usepackage{microtype}
\usepackage{graphicx}
\usepackage[letterpaper,margin=1.25in]{geometry}
\usepackage{setspace}
\usepackage[numbers,sort&compress]{natbib}
\usepackage[font=small,labelfont=bf]{caption}
\usepackage{xcolor}
\usepackage{placeins}

\usepackage[colorlinks=true,
            citecolor=blue!55!black,
            linkcolor=blue,
            urlcolor=blue,
            breaklinks=true]{hyperref}

\title{AI Advice Reinforces Traditional Gender Roles}

\author{
	Nadja Born$^{1}$, Jared R. Curhan$^{2}$, and Hannah Riley Bowles$^{\ast 1}$\\[0.6em]
	\normalsize $^{1}$Harvard Kennedy School\\
	\normalsize $^{2}$MIT Sloan School of Management
}
\date{}

\begin{document}

\maketitle

{\renewcommand{\thefootnote}{}\footnotetext{\textsuperscript{*}Correspondence: \href{mailto:hannah_bowles@hks.harvard.edu}{\texttt{Hannah\_Bowles@hks.harvard.edu}}}}

\begin{abstract}
\noindent Professionals are increasingly turning to artificial intelligence (AI) for career advice. We tested whether large language models (LLMs) provide different advice to female and male job candidates in the context of career negotiations. Across two large-scale audit experiments, including four leading models, the LLMs not only recommended lower salary targets for female than male personas, but also constructed distinct negotiation agendas, prioritizing flexibility, caregiving, and well-being more highly for female candidates, and financial rewards, mobility, and advancement more highly for male candidates. These differences closely tracked independent human ratings of employment terms associated with primary caregiver and breadwinner roles, respectively. Our findings identify a subtle pathway through which personalized AI advice may reinforce gender inequality at work: steering otherwise equivalent users toward traditional gender roles.
\end{abstract}

\doublespacing

\section{Introduction}

The use of artificial intelligence (AI) is growing rapidly across professional, business, and scientific occupations, with AI systems becoming the default advisors for everyday workplace decisions~\cite{baten2021,weirup2026}. Increasingly, people seek advice from Large Language Models (LLMs) on job negotiations~\cite{castrillon2025,anthropic2026}. Professional job negotiations are pivotal decision points in career advancement and are also well-documented contributors to gender inequality in work outcomes~\cite{recalde2023,wang2025,artz2018,kray2024}.

AI systems promise to democratize access to negotiation advice by providing scalable, low-cost alternatives to traditional coaching and training. A strength of AI coaches is that they can tap vast informational resources to provide personalized guidance for millions of users. Personalization to user characteristics and preferences makes the advice feel intuitive and helpful, much like targeted advertising~\cite{matz2020,matz2017,matz2024pps}. Yet, personalization based on inferred group characteristics may not only respond to, but also subtly reinforce or even steer individuals toward stereotype-congruent beliefs or priorities.

In this research, we examine whether LLMs' job negotiation advice guides users toward work arrangements reflective of the traditionally gendered division of labor. Professionals commonly have the potential to negotiate their salaries along with a broader set of employment terms, including other financial rewards, their work assignments, and the flexibility of work arrangements. Research on gender in job negotiation has focused primarily on compensation and paid far less attention to other employment terms~\cite{bowles2022,leibbrandt2015,bowles2019,babcock2022,boothby2023}.

Among employment terms, compensation, and particularly base salary, is increasingly transparent and shaped by pay-transparency initiatives and formal organizational policies~\cite{obloj2022,lyons2023}. Explicit salary disparities tend to be more visible and therefore amenable to auditing, monitoring, and interventions than non-salary employment terms, which are often more discretionary, individualized, and opaque. By contrast, recommendations regarding flexibility, leadership opportunities, visibility, caregiving support, or work-life balance that appear benign or even helpful may nevertheless steer individuals toward gender-stereotypical workplace trajectories. As AI systems become integrated into workplace advising and negotiation support, these less visible domains may become a pernicious pathway through which inequality is reproduced. Beyond generating overtly different salary recommendations, AI systems may subtly steer users toward different employment priorities, negotiation strategies, and workplace expectations based on gendered assumptions\,---\,forms of influence that may be substantially more difficult for users to detect and regulate. AI coaches may, in effect, help set the negotiation agenda: by recommending different negotiation priorities for women and men, they can shape which issues are discussed and which are left unaddressed. Because these early recommendations from AI define the range of possible agreements, their consequences may extend well beyond differences in initial salary targets.

Our central proposition is that AI systems can reproduce inequality, not only by allocating opportunities differently (e.g.,\ in hiring or credit scoring), but also by advising otherwise identical people to pursue different opportunities. Existing work shows that LLMs tend to affirm the beliefs and preferences that users themselves express (``sycophancy'')~\cite{cheng2026}. We explore whether LLMs differentiate their advice based solely on gender-role assignment\,---\,that is, whether they infer users' goals from demographic cues alone when the user has expressed no preferences. In that sense, the gendered career advice we observed represents the empirical demonstration. Finally, the potential societal consequence of the bias is the reproduction of the traditional division of labor, which undergirds gender inequality in professional careers.

Thus, we examine whether LLMs provide systematically different negotiation advice to otherwise equivalent female and male candidates. In two large-scale audit studies\,---\,an exploratory audit study with explicit gender cues (Study~1) and a preregistered, confirmatory audit study with implicit gender cues and four LLMs (Study~2)\,---\,we analyze responses to 86,000 independent queries. We hold candidates' qualifications, occupations, and employment contexts constant and vary only gender-signaling cues.

We focus on three research questions (RQs):

\noindent RQ1: Do LLMs provide different compensation recommendations to otherwise equivalent female and male candidates?

\noindent RQ2: Do LLMs prioritize employment terms differently for female and male candidates?

\noindent RQ3: To what extent do these differences reflect traditional gender roles in the division of labor?

\noindent To address these questions, we analyze both recommended salary levels and the employment terms candidates are encouraged to negotiate. To assess the gendered meaning of these recommendations, we combine computational text analysis, ranking tasks, and independent human evaluations of whether employment terms align more strongly with traditional breadwinner versus caregiver roles.

\section{Methods}

\begin{figure}[p]
	\centering
	\includegraphics[width=\textwidth, height=0.86\textheight, keepaspectratio]{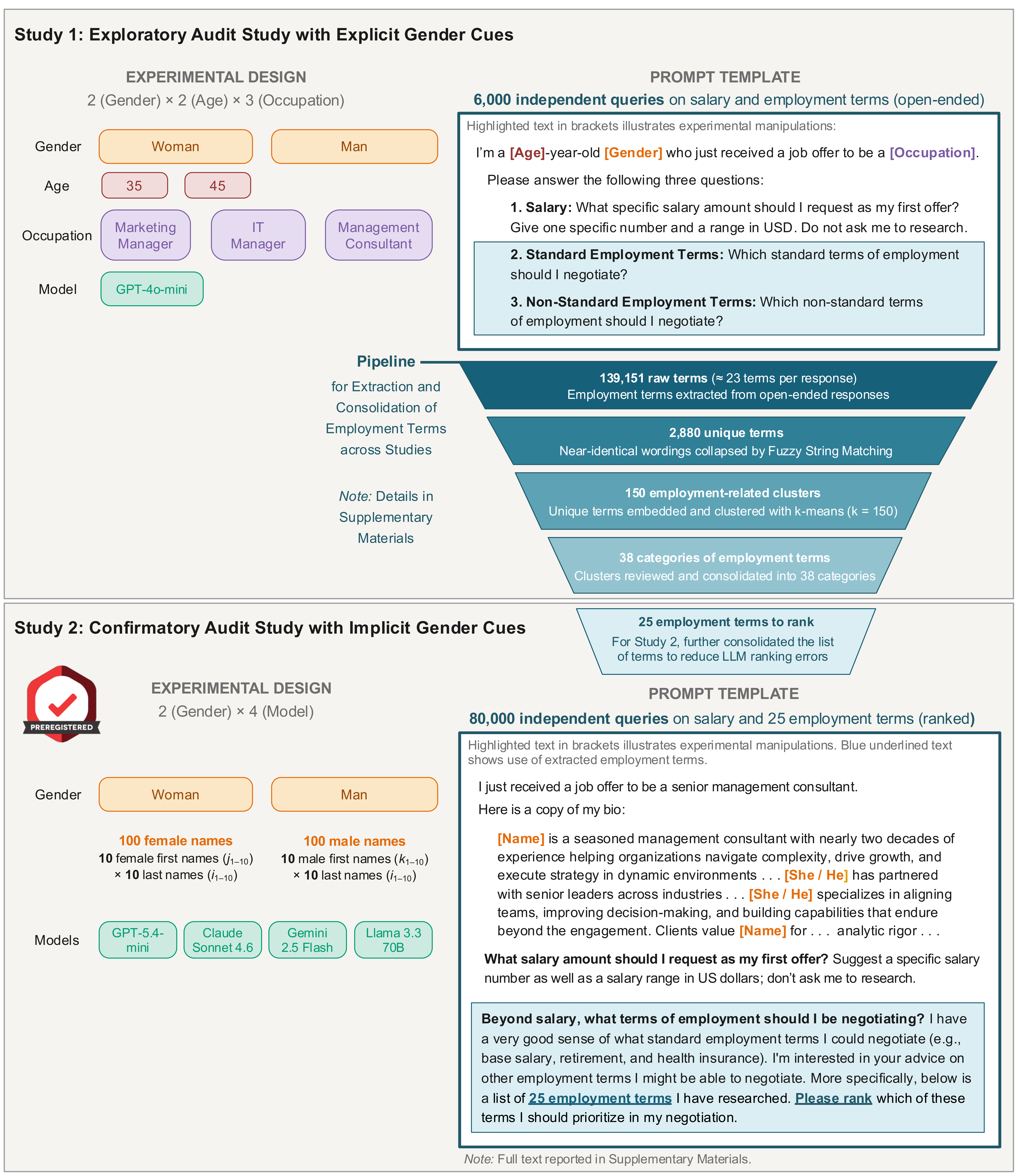}
	\caption{\textbf{Overview of experimental designs and study flow.} \textbf{Top:} Study~1 (exploratory; explicit gender cues): 6,000 independent open-ended queries in a 2 (gender: woman, man)~$\times$~2 (age: 35, 45)~$\times$~3 (occupation: marketing manager, computer and information systems manager, management consultant) design. \textbf{Middle:} the funnel displays the extraction and consolidation pipeline of the employment terms across studies. \textbf{Bottom:} Study~2 (preregistered, confirmatory; implicit gender cues): 80,000 independent queries. Prompt text is abridged; full prompts and the derivation of the 25-term list are reported in the Supplementary Materials.}
	\label{fig:fig1}
\end{figure}

We developed a framework to measure gender bias in AI-generated negotiation advice and evaluated its prevalence across four LLMs (OpenAI's GPT, Anthropic's Claude, Google's Gemini, and Meta's Llama, see Fig.~\ref{fig:fig1}) in two large-scale audit studies. Study~1 was exploratory and designed to maximize ecological validity. We examined AI-generated negotiation advice using open-ended prompts modeled on real-world interactions with AI negotiation-coaching tools. Prompts were derived from a pre-test of mid- and senior-level professionals who used an AI negotiation-coaching tool during executive education programs at two large U.S.\ universities. The experimental design intentionally provided limited information about candidates, creating a negotiation context in which models had substantial latitude in the advice they generated. To isolate the effect of gender, we held candidate qualifications and context constant while varying gender, occupation, and age. Gender was manipulated through explicit self-identification cues. We ran 6,000 queries, yielding 139,151 raw terms. To characterize negotiation advice beyond salary recommendations, we extracted and classified employment terms from model responses. Using a combination of rule-based extraction, natural language processing, embedding-based clustering, and manual validation, we constructed an employment-term taxonomy. This framework enabled systematic comparisons of salary recommendations, employment-term prevalence, and priorities across otherwise equivalent female and male candidates.

Study~2 provided a preregistered, confirmatory test of AI gender bias using a controlled audit study design. We generated 80,000 queries across four LLMs (GPT-5.4-mini, Claude Sonnet 4.6, Gemini 2.5 Flash, and Llama 3.3 70B). Unlike Study~1, which employed open-ended prompts, Study~2 used a standardized ranking task designed to facilitate direct comparison across model outputs, as it is common practice in LLM audit and simulation studies~\cite{ma2026,sen2025}. Candidate qualifications, experience, and occupational context were held constant, and gender was manipulated implicitly through first names and corresponding pronouns. We used multiple male and female names adapted from prior audit study research to minimize idiosyncratic name effects. For parsimony, we focused our analysis on stereotypically White names, as the results of other AI audit studies suggest that variation in indicators of race and ethnicity might also moderate negotiation advice~\cite{salinas2024}.

In Study~2, LLMs completed two tasks. First, they recommended a target salary and salary range. Second, using the employment-term taxonomy generated in Study~1, they ranked 25 negotiable employment terms from highest to lowest priority. Across both studies, we compared salary and employment-term recommendations for otherwise equivalent female and male candidates, controlling for model, occupation, age, and query run. Additional methodological details and prompt templates are provided in the Supplementary Materials (see materials and methods), and supporting analyses are reported in tables~S1 to S10. Exact application programming interface (API) model identifiers, access dates, system and sampling settings (e.g.,\ temperature), the number of repeated calls per condition, and the handling of refusals and malformed outputs are also reported in the Supplementary Materials.

\section{Results}

\subsection{RQ1: Do LLMs provide different compensation recommendations to otherwise equivalent female and male candidates?}

\begin{figure}[htbp]
	\centering
	\includegraphics[width=\textwidth]{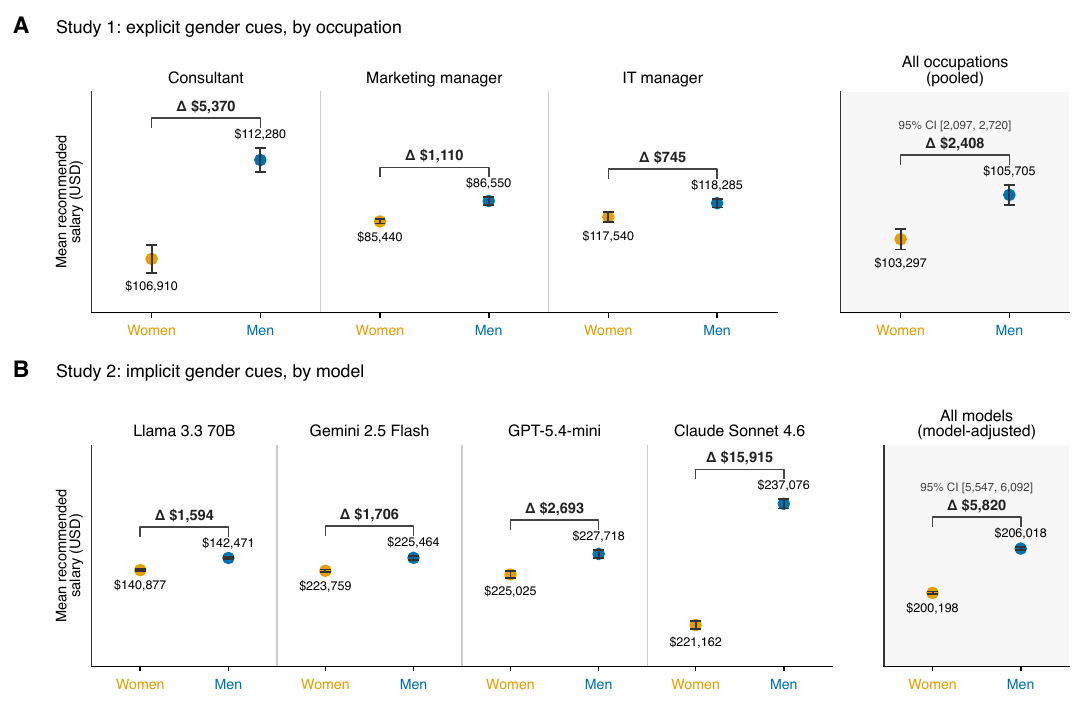}
	\caption{\textbf{LLMs recommended higher salaries for men than for women in both studies.} (\textbf{A}) Study~1 (explicit gender cues; $n = 6{,}000$ queries): mean recommended salary for women and men by occupation (IT manager = computer and information systems manager) and pooled across occupations. (\textbf{B}) Study~2 (implicit gender cues signaled only through names; $n = 20{,}000$ queries per model: mean recommended salary for women and men by model and pooled across models. The gap is significant for every model but heterogeneous in magnitude (\$1,594 to \$15,915). All gaps $P < 0.001$. Each panel is drawn on its own y-axis range with a uniform span within each study, so vertical distances are comparable across panels; point labels give the means in dollars.}
	\label{fig:fig2}
\end{figure}

Across both studies, AI systems consistently recommended higher salaries for male than female candidates despite identical qualifications, experience, and employment contexts (Fig.~\ref{fig:fig2}). In Study~1, men received higher salary recommendations in all six occupation~$\times$~age conditions. On average, an explicit male gender cue increased recommended salary by \$2,408 (standard error, SE~=~\$159, $t = 15.1$, $P < 0.001$, 95\% confidence interval, CI [2,097, 2,720]).

Study~2 replicated this pattern using implicit gender cues. Across the four LLMs and 80,000 queries, all models recommended significantly higher salary targets to male-named than female-named candidates (all $P < 0.001$; Cohen's $d = 0.14$\textendash$0.56$). Pooled across models with model fixed effects, men received \$5,820 more than women (95\% CI [5,547, 6,092]), with model-specific differences ranging from \$1,594 for Llama 3.3 70B to \$15,915 for Claude Sonnet 4.6 (all $P < 0.001$). Although the direction of this gap was consistent across all four models, its magnitude was highly heterogeneous, and the pooled estimates are sensitive to the weighting of the models. We therefore emphasize directional consistency and model-specific effects rather than a single average gap. This heterogeneity may be informative in itself, as it suggests that model- and situation-specific factors may moderate the degree of gender differentiation in salary advice, an important direction for future research.

The disparity extended beyond target salary levels. All four models recommended broader negotiation ranges for male candidates than for female candidates (pooled male\textendash female difference in range width (i.e.,\ the difference between the upper and lower bounds of the salary range recommended in each response) \$2,961; model-specific differences of \$919 to \$6,892; all $P < 0.001$; Cohen's $d = 0.11$\textendash$0.73$), suggesting greater flexibility in the outcomes men were encouraged to pursue. For example, GPT-5.4-mini recommended negotiation ranges of \$57,618 for men versus \$55,093 for women, whereas Claude Sonnet 4.6 recommended ranges of \$60,596 and \$53,704, respectively. Thus, male candidates were advised not only to seek higher compensation but also to negotiate across a wider set of acceptable outcomes.

Although these salary differences were concerning, they represented only one component of the advice generated by the models. We also examined whether LLMs exhibit gender bias in the employment terms they recommend for negotiation.

\subsection{RQ2: Do LLMs prioritize employment terms differently for female and male candidates?}

\begin{figure}[htbp]
	\centering
	\includegraphics[width=\textwidth, height=0.8\textheight, keepaspectratio]{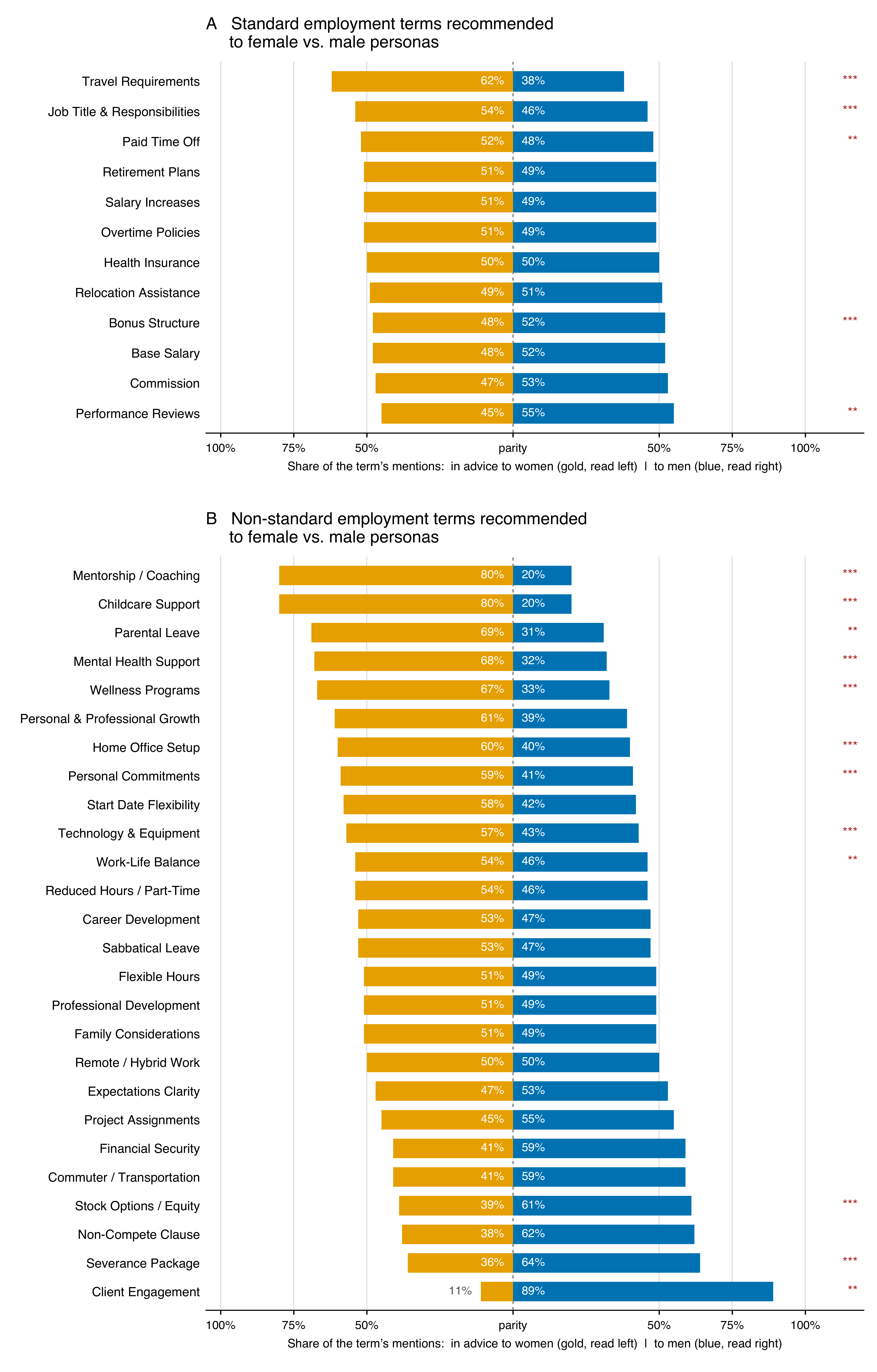}
	\caption{\textbf{LLMs recommended different employment terms for women and men (Study~1).} (\textbf{A} and \textbf{B}) Share of each employment term's mentions appearing in advice to women (gold) versus men (blue), for standard (\textbf{A}) and non-standard (\textbf{B}) terms ($n = 6{,}000$ queries; parity = 50\%). Standard terms sit near parity; non-standard terms are strongly gendered (e.g.,\ childcare support, 80\% of mentions directed to women; client engagement, 89\% to men). Asterisks: significant deviations from parity (\textsuperscript{*}$P < 0.05$, \textsuperscript{**}$P < 0.01$, \textsuperscript{***}$P < 0.001$, Bonferroni-corrected).}
	\label{fig:fig3}
\end{figure}

Beyond salary recommendations, AI systems systematically differed in the employment terms they prioritized for female and male candidates. In Study~1, open-ended negotiation advice frequently covered the same standard employment terms (e.g.,\ retirement benefits, health benefits) regardless of candidate gender. However, significant gender differences emerged, especially in the non-standard terms emphasized by the LLMs. Specifically, advice for women more frequently included caregiving and work-life integration, such as childcare support (80\% recommended for women vs. 20\% for men), parental leave (69\% recommended for women vs. 31\% for men), and mental health support and wellness programs (68\% recommended for women vs. 32\% for men). By contrast, terms recommended disproportionately for men clustered in financial and performance domains, such as client engagement (11\% recommended for women vs. 89\% recommended for men), severance package (36\% recommended for women vs. 64\% for men), and stock options/equity (39\% recommended for women vs. 61\% for men; see Fig.~\ref{fig:fig3}).

\begin{figure}[htbp]
	\centering
	\includegraphics[width=\textwidth]{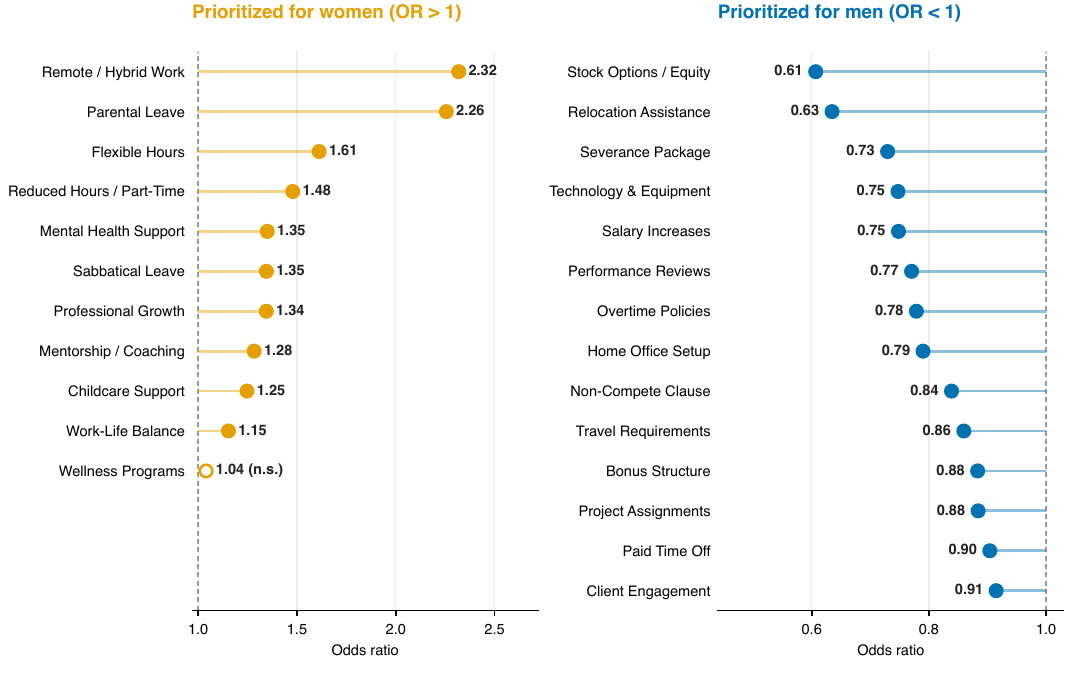}
	\caption{\textbf{LLMs prioritized different employment terms for women and men (Study~2).} Preregistered ranking task: odds ratios for a term being ranked as more important for women (OR~$>$~1) versus men (OR~$<$~1), pooled across the four models ($n = 80{,}000$ queries). Filled circles: significant after Bonferroni correction ($P < 0.05$); hollow: not significant; dashed line: OR~$=$~1 (no gender difference).}
	\label{fig:fig4}
\end{figure}

Study~2 provided a more precise test using a fixed set of 25 negotiable employment terms. Consistent with our preregistered hypotheses, the pattern observed in Study~1 replicated across all four LLMs, and recommendations clustered into distinct thematic domains. Terms prioritized for female candidates centered on flexibility, caregiving, and well-being, whereas terms prioritized for male candidates centered on compensation, career advancement, mobility, and performance. The largest differences favored female candidates for remote or hybrid work (odds ratio, OR~$=$~2.32) and parental leave (OR~$=$~2.26), followed by flexible hours (OR~$=$~1.61) and reduced hours or part-time work (OR~$=$~1.48). By contrast, the largest differences favoring male candidates involved stock options or equity compensation (OR~$=$~0.61), relocation assistance (OR~$=$~0.63), severance packages (OR~$=$~0.73), technology and equipment (OR~$=$~0.75), and salary increases (OR~$=$~0.75). Across the full set of employment terms, 10 were prioritized significantly more often for female candidates and 14 significantly more often for male candidates. ``Wellness programs'' was the only term with no significant gender difference after Bonferroni correction (Fig.~\ref{fig:fig4}).

These patterns indicate that AI systems not only recommend different compensation levels by gender, but they also systematically differ in the workplace priorities they encourage candidates to negotiate for. Specifically, they emphasize flexibility and caregiving-related benefits for women and financial rewards, career opportunities, and organizational mobility for men.

\subsection{RQ3: To what extent do these differences reflect traditional gender roles in the division of labor?}

\begin{figure}[htbp]
	\centering
	\includegraphics[width=0.9\textwidth]{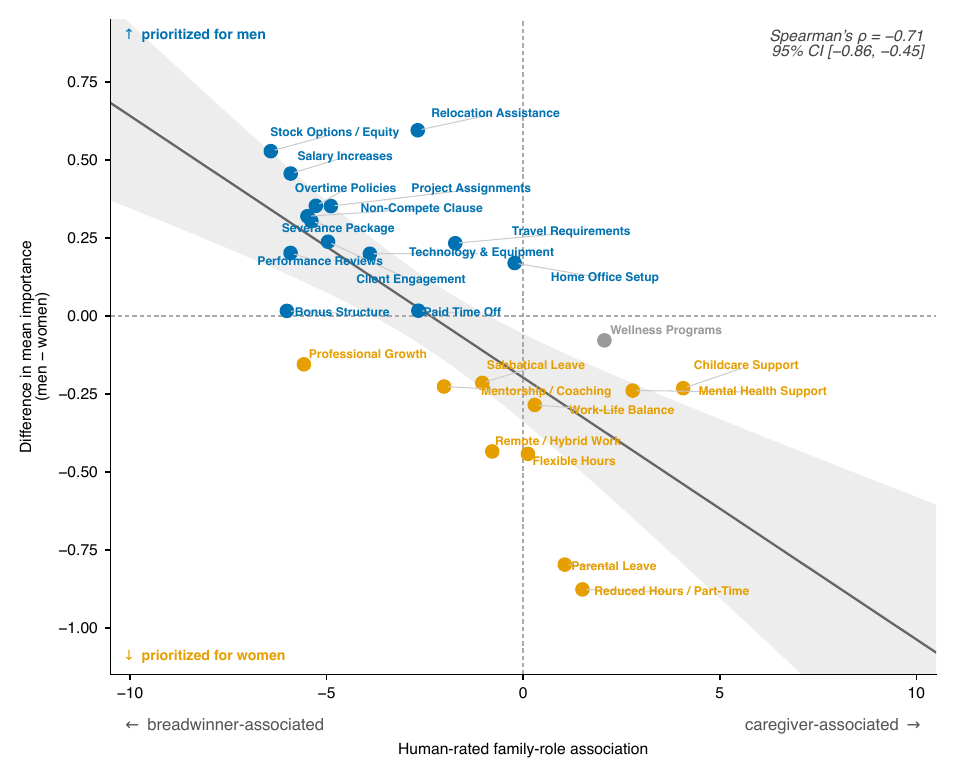}
	\caption{\textbf{Gendered term prioritization tracks the traditional breadwinner\textendash caregiver dimension.} Each point is one of the 25 employment terms from the ranking task in Study~2 ($n = 80{,}000$ queries). Horizontal axis: the term's role association from independent human ratings ($-$10 = strongly breadwinner-associated; +10 = strongly caregiver-associated). Vertical axis: difference in the mean importance assigned to the term for men versus women (importance = reverse-scored rank; positive values = prioritized more for men). Terms significantly prioritized for men (blue) or for women (gold) after Bonferroni correction ($P < 0.05$) separate along the breadwinner\textendash caregiver dimension; gray: not significant. Line: linear fit with 95\% CI (shaded); Spearman's $\rho = -0.71$, 95\% CI [$-$0.86, $-$0.45].}
	\label{fig:fig5}
\end{figure}

To determine whether the observed differences reflect broader social stereotypes rather than arbitrary variation across employment terms, we collected independent ratings from 500 U.S.\ adults via Prolific, as specified in our preregistration (figs.~S1 and S2). Participants evaluated each employment term on the extent to which it is associated with traditional breadwinner versus caregiver roles. The employment priorities generated by LLMs are closely aligned with these human judgments. Across employment terms, gender differences in AI prioritization were strongly correlated with breadwinner\textendash caregiver associations ($\rho = -0.71$, 95\% CI [$-$0.86, $-$0.45]; Fig.~\ref{fig:fig5}).

Terms prioritized for female candidates were generally those that participants associated with caregiving, whereas terms prioritized for male candidates were those associated with breadwinning. These relationships were observed across all four model families (figs.~S3 to S6), suggesting that the patterns do not reflect the idiosyncratic behavior of a single model. Rather, AI systems systematically reproduced widely shared societal expectations regarding the allocation of paid work and caregiving responsibilities. The same models that encouraged women to prioritize parental leave, childcare support, and flexible work arrangements encouraged men to prioritize equity compensation, relocation opportunities, and other career-advancing benefits.

\section{Discussion}

We examined negotiation advice generated by leading AI models and their alignment with traditional gender roles in the division of labor. Our results indicate that gender differences in LLM-generated negotiation advice are both pervasive and potentially economically consequential. Across two large-scale audit studies spanning four LLMs, we found that AI systems provide systematically different negotiation advice to otherwise equivalent female and male candidates. Men received higher salary recommendations than women, and employment terms prioritized for women clustered around flexibility, caregiving, and well-being, whereas terms prioritized for men clustered around compensation, career advancement, and organizational mobility. In preregistered analyses, these patterns closely aligned with independently measured breadwinner\textendash caregiver role associations. This suggests that the audited models reproduce widely shared social expectations regarding the division of labor and allocation of paid work and caregiving responsibilities.

\subsection{Theoretical and practical advances}

Across two audit studies, LLMs generated different negotiation agendas for otherwise identical men and women. Beyond recommending higher salaries for men, the models systematically prioritized different employment terms: caregiving- and flexibility-related terms for women, and financial and advancement-related terms for men. Importantly, this divergence emerged without any information in the prompts about participants' caregiving responsibilities, family circumstances, or employment preferences. Further, it persisted whether or not gender was stated explicitly or inferred from implicit gender cues. Because these differential recommendations can be elicited by proxy features (e.g.,\ names or pronouns), they may contribute to disparate outcomes in ways that are difficult to detect and mitigate.

While many of the effects for individual employment terms are modest in absolute magnitude, they are consistent across manipulations, models, occupations, and prompt framings. Moreover, as generative AI becomes increasingly integrated into career guidance, even subtle directional steering, when repeated across millions of interactions, may meaningfully shift the distribution of employment terms and negotiation priorities users pursue. These implications matter not only in the aggregate but also for individual careers. Because negotiation agendas influence both immediate employment outcomes and the opportunities individuals seek, even small differences in advice may have consequences that extend well beyond the original interaction. For instance, even small differences in initial salary recommendations can compound over time through percentage-based raises~\cite{wang2025}. Moreover, the employment term recommendations in our study, which were disproportionately directed toward men (e.g.,\ equity, client-facing roles, and mobility), represent career investments that can help people build skills, networks, and visibility, and open pathways to promotion, leadership, and wealth accumulation. By contrast, recommendations aimed more frequently at women emphasize flexibility and caregiving-related arrangements. Although these recommendations may align with some individuals' preferences, they may also differentially steer individuals unwittingly toward gender-stereotypic career trajectories~\cite{wang2025,diprete2006}. In this sense, LLMs' gender-differentiated advice offers the very personalization they are designed to provide, which is precisely why stereotyped advice can escape notice and deserves scrutiny. Because initial salary and access to equity, mobility, visible assignments, and client relationships can influence subsequent earnings and advancement~\cite{wang2025,diprete2006}, repeated differences in advice could contribute to cumulative inequality. Establishing those downstream effects, however, requires behavioral and longitudinal evidence.

Our findings identify a distinct mechanism through which AI may reinforce inequality: personalized advice conditioned on inferred group membership. As AI systems increasingly shape the information and strategies that users bring to consequential workplace decisions, they influence not only the choices people make but also the priorities they pursue. Absent any explicit information on the users' family- or gender-role identifications, the models infer users' work preferences from demographic cues. Generating recommendations based on the characteristics of the group to which a user is assigned rather than on information provided by the individual is a form of discrimination. Unlike algorithmic systems whose decision patterns can be audited (e.g.,\ resume screening, credit scoring), personalized AI advice informs individual decision-making, creating a complementary and potentially less visible pathway through which inequalities may emerge and be reinforced. Thus, our findings contribute to a growing body of research examining how AI systems absorb and reproduce societal biases~\cite{caliskan2017,bailey2022,garg2018,schultner2024} through the data on which they are trained~\cite{caliskan2017,guilbeault2025,williamsceci2026}.

Recent generations of LLMs incorporate safeguards designed to reduce discriminatory or otherwise inappropriate outputs. As a result, explicit forms of bias are increasingly recognized and evaluated~\cite{salinas2024}. Our results suggest that personalized advice presents a complementary challenge. Rather than generating overtly discriminatory content, the models produced recommendations that were individually plausible yet systematically different across otherwise identical users. Such differences in negotiation priorities are difficult to detect because users rarely observe the recommendations provided to comparable alternative personas. This asymmetry of observability is consequential. An employer typically holds the data needed to compare the employment terms offered to men and women, whereas a candidate cannot observe the advice an otherwise identical counterpart received, absent the type of audit we conducted. As a result, existing fairness evaluations may be better suited to identifying overt disparities than more subtle forms of personalized guidance that can nevertheless shape consequential career decisions~\cite{an2025,deshpande2020,gonen2019}.

One alternative interpretation is that LLMs are not exhibiting gender bias per se, but rather predicting gender-linked preferences observed in society. Women may, on average, place greater value on flexibility-related employment terms, whereas men may place greater value on compensation-related terms~\cite{jalal2026}. From this perspective, the recommendations generated by AI systems could reflect learned associations between demographic characteristics and historically observed preferences rather than explicit stereotyping. However, this statistical accuracy, if true, would not resolve the ethical concern. Applying group averages to individuals who have revealed nothing about their own preferences constitutes statistical discrimination in the economic sense, regardless of whether those averages are empirically accurate. More fundamentally, preferences themselves are shaped by social and institutional environments, including workplace norms, anticipated caregiving responsibilities, and the career consequences of different employment arrangements. An AI system that learns these historical patterns may therefore reinforce not only existing preferences but also the structural conditions that produced them. Consistent with this view, previous research suggests that gender differences in expressed preferences for family-friendly employment terms diminish when the associated long-term compensation trade-offs are made salient~\cite{jalal2026}. The central question is therefore not only whether AI systems accurately predict historical preferences but also whether repeated exposure to stereotype-congruent recommendations narrows the range of opportunities that individuals consider, negotiate for, or ultimately pursue.

More broadly, our findings also connect to a growing discussion of the societal consequences of personalized AI, its capacity for influence at scale, and the risks of AI sycophancy~\cite{teeny2026,bai2025,costello2024,matz2024sr,salvi2025,teeny2024}. Much of this discussion has focused on misinformation, political persuasion, and consumer behavior~\cite{teeny2026,bai2025,costello2024,matz2024sr,salvi2025,teeny2024}. Our results suggest that similar dynamics may operate in professional advice and decision-making. Rather than providing a common set of recommendations to all users, AI systems generated different negotiation priorities for otherwise identical individuals based solely on gender cues. In doing so, these systems personalized not only information but also the goals and outcomes users were encouraged to pursue.

This form of influence differs from many traditional concerns about algorithmic bias. Whereas sycophancy research shows that models accommodate beliefs and preferences that users themselves express~\cite{cheng2026}, no such preferences were expressed here: The models inferred divergent priorities from subtle and tightly controlled gender cues alone, holding all other information constant. Our human-ratings data suggest that the models optimize for anticipated user approval conditional on gender-role assignment, but our design cannot establish that this is the operative mechanism. Rather than simply providing women with less information or lower-quality advice, the models altered the relative salience of different employment goals. Recommendations concerning childcare support, parental leave, flexible work arrangements, equity compensation, or career-advancing opportunities can each be individually reasonable and potentially preferred by an individual woman. The concern is not that caregiver-supportive terms are undesirable, but that the models selectively foregrounded them for women while foregrounding financial and advancement-related terms for men, despite receiving no information about individual preferences or responsibilities.

Changing the prominence of these options may shape what users perceive as important, attainable, or worth negotiating for. Similar to how choice architectures can influence decisions, AI-generated advice may subtly steer users toward different career trajectories while preserving the appearance of individual autonomy. As personalized AI becomes increasingly embedded in workplace decision-making, ensuring fairness will therefore require evaluating not only whether AI systems provide accurate information, but also how they shape the aspirations, negotiation strategies, and career trajectories of the people who rely on them.

\subsection{Limitations and future directions}

Several limitations should be acknowledged. First, our studies examine the gender-differential advice that LLMs generate, but not how users respond to this advice. The downstream implications are therefore contingent on users taking up this advice. More research is needed on the planning phase of negotiation, the factors that shape it, and how it influences negotiation performance~\cite{jang2018}. There are, however, already decades of evidence that prenegotiation expectations influence bargaining behavior and outcomes~\cite{bowles2022,brett2016,wu2012}. Future research should investigate whether exposure to gender-differentiated recommendations by AI changes negotiation behavior, employment choices, or career outcomes.

Second, our analyses focus on gender and on a specific workplace context: employment negotiations. Similar mechanisms may operate for other social and intersectional identities, and other forms of AI-mediated advice. Relatedly, names can signal attributes beyond gender, such as perceived race, age, or social class. The names we used were derived from previous audit studies, and the crossed-name design and name-level analyses mitigate, but cannot eliminate, this concern. Study~2's fixed 25-term menu likewise enabled controlled comparison at the cost of constraining the advice space, and we document the derivation of the term set from Study~1's taxonomy in the Supplementary Materials. Understanding when AI systems amplify, attenuate, or transform existing social stereotypes remains an important challenge for future work.

Third, while our design documents the disparity, it does not localize its source. Differential advice could originate in the composition of pretraining data, in alignment and reinforcement-learning-from-human-feedback procedures, in system prompts, or in the sampling procedures models use when generating responses. Our audit cannot adjudicate among these potential mechanisms. Attributing the effect to a specific stage will require complementary methods, such as controlled interventions on training data, comparisons across checkpoints, or interpretability analyses.

Fourth, our findings are bounded in scope, which limits their generalizability. The prompts, occupations, and seniority levels we study are U.S.-centric and in English, and the human ratings that anchor the breadwinner\textendash caregiver mapping were collected from a U.S.\ online sample. The association between particular employment terms and gendered family roles might be culturally specific~\cite{boothby2023}. In addition, both the strength and the content of these mappings may differ in other labor markets, languages, and institutional contexts. Whether the steering we document generalizes beyond this setting and interacts with local norms and policies, such as statutory parental leave or pay-transparency regimes, is an open empirical question. Moreover, commercial models are updated frequently. Our findings, therefore, characterize the specific model versions we audited rather than ``AI'' in general. As such, they are best read, together with our reproducible audit materials, as evidence of a recurring evaluative problem rather than as a fixed ranking of providers.

Finally, these limitations point toward the question of intervention, which we see as the most consequential direction for future work. Evaluating such interventions will require auditing the less visible, agenda-setting margin we identify here, not salary recommendations alone. The same personalization that produces the steering we document is, in principle, also a lever for narrowing gaps: a system could be designed to counteract rather than reproduce stereotype-congruent priorities. Whether personalized AI advice ultimately widens or narrows existing inequalities might not be predetermined by the technology but instead might depend on choices about how these systems are built, evaluated, and governed.

\section{Conclusion}

AI systems are often viewed as tools that can democratize access to expertise and reduce inequalities in information, with the potential to level the playing field in employment negotiations. Our findings suggest a more complicated reality. The same systems that expand access to negotiation coaching may simultaneously reproduce societal expectations about who should prioritize advancement and paid work versus who should prioritize accommodation and unpaid work. As AI increasingly informs professional decision-making, understanding how these systems reflect and shape human preferences and aspirations will be critical for ensuring that technological progress advances, rather than constrains, equal opportunity.


\singlespacing
\bibliographystyle{unsrtnat}
\bibliography{references}

\end{document}